\documentclass[aps,prd,preprint,superscriptaddress,showpacs]{revtex4}
\usepackage{graphicx}
\usepackage{verbatim}
\usepackage{ulem}
\usepackage{color}
\definecolor{My_red}        {cmyk}{0.00,1.00,1.00,0.20}

\newcommand{\bmat}{\left(\begin{array}}
\newcommand{\emat}{\end{array}\right)}
\newcommand{\beq}{\begin{equation}}
\newcommand{\eeq}{\end{equation}}
\newcommand{\wt}{\widetilde}

\def\ra{\rightarrow}

\def\ld{\lambda}
\def\f{\frac}

\def\bwt{\begin{widetext}}
\def\ewt{\end{widetext}}
\def\be{\begin{equation}}
\def\ee{\end{equation}}
\def\bea{\begin{align}}
\def\eea{\end{align}}
\def\bean{\begin{align*}}
\def\eean{\end{align*}}
\def\bary{\begin{array}}
\def\eary{\end{array}}
\def\bit{\begin{itemize}}
\def\eit{\end{itemize}}

\def\ra{\rightarrow}

\def\ld{\lambda}

\def\su5u1{SU(5) \times U(1)}
\def\fsu5u1{SU(5) \times U(1)'}
\def\so10{SO(10)}
\def\sq20{SO(10) \times SO(10)}

\def\ra{\rightarrow}

\def\ld{\lambda}
\def\f{\frac}

\def\L{\left(}
\def\R{\right)}

\def\ra{\rightarrow}

\def\ld{\lambda}

\def\su5u1{SU(5) \times U(1)}
\def\fsu5u1{SU(5) \times U(1)'}
\def\so10{SO(10)}
\def\sq20{SO(10) \times SO(10)}

\usepackage[centertags]{amsmath}
\usepackage{amssymb}

\begin{document}

\title{Higgs Boson Mass and Complex Snuetrino Dark Matter \\
in the Supersymmetric Inverse Seesaw Models}

\author{Jun Guo}
\email{hustgj@itp.ac.cn}

\affiliation{State Key Laboratory of Theoretical Physics
and Kavli Institute for Theoretical Physics China (KITPC),
      Institute of Theoretical Physics, Chinese Academy of Sciences,
Beijing 100190, P. R. China}

\author{Zhaofeng Kang}
\email{zhaofengkang@gmail.com}

\affiliation{Center for High-Energy
Physics, Peking University, Beijing, 100871, P. R. China}

\author{Tianjun Li}
\email{tli@itp.ac.cn}

\affiliation{State Key Laboratory of Theoretical Physics
and Kavli Institute for Theoretical Physics China (KITPC),
      Institute of Theoretical Physics, Chinese Academy of Sciences,
Beijing 100190, P. R. China}

\affiliation{School of Physical Electronics,
University of Electronic Science and Technology of China,
Chengdu 610054, P. R. China}

\author{Yandong Liu}
\email{ydliu@itp.ac.cn}

\affiliation{State Key Laboratory of Theoretical Physics
and Kavli Institute for Theoretical Physics China (KITPC),
      Institute of Theoretical Physics, Chinese Academy of Sciences,
Beijing 100190, P. R. China}

\date{\today}

\begin{abstract}

The discovery of a relatively heavy Standard Model (SM)-like Higgs
boson challenges naturalness of the minimal supersymmetric standard
model (MSSM) from both Higgs and dark matter (DM) sectors. We study
these two aspects in the MSSM extended by the low-scale inverse
seesaw mechanism. Firstly, it admits a sizable radiative contribution
to the Higgs boson mass $m_h$, up to $\sim$ 4 GeV in the case of an
IR-fixed point of the coupling $Y_\nu LH_u\nu^c$ and a large
sneutrino mixing. Secondly, the lightest sneutrino, highly complex
as expected, is a viable thermal DM candidate. Owing to the correct
DM relic density and the XENON100 experimental constraints, 
two scenarios survive:  a Higgs-portal
complex DM with mass lying around the Higgs pole or above $W$
threshold, and a coannihilating DM with slim prospect of detection.
Given an extra family of sneutrinos, both scenarios naturally work
when we attempt to suppress the DM left-handed sneutrino component,
confronting with enhancing $m_h$.

\end{abstract}

\pacs{12.60.Jv, 14.70.Pw, 95.35.+d}

\maketitle

\section{Introduction and motivations}

The CMS and ATLAS Collaborations discovered a new resonance 
around 125.5 GeV~\cite{LHC:Higgs}. From the latest full collected
data announced at the Moriond 2013 conference, it is quite Standard
Model (SM)-like. If this is confirmed at the next run of the
$\sqrt{s}=14$ TeV LHC, it would complete the picture of SM.
But we can never conclude that the discovery of a highly SM-like
Higgs boson at the LHC indicates an end to the particle physics: on
the theoretical side, the SM suffers the notorious gauge hierarchy
problem if the discovered resonance is indeed a fundamental spin-0
boson; On the phenomenological side, the SM can not explain the tiny
neutrino mass origin and has no candidate for dark matter (DM), both
of which are clear signals for new physics beyond the SM.

Supersymmetry (SUSY) is still the most promising underlying theory
to account for these two sides simultaneously. The supersymmetric
SMs (SSMs) are free of quadratic divergences involving scalars, and
provide a weakly interactive massive particle (WIMP) DM
candidate if $R$-pariy is conserved, {\it i.e.}, the Lightest
Supersymmetric Particle (LSP) such as the lightest
neutralino~\cite{SUSYDM}. Of course, to explain the neutrino masses
and mixings, we may have to supersymmetrize the well
studied models with seesaw mechanisms. Among them, the inverse
seesaw (ISS) mechanism~\cite{Mohapatra:1986bd} has an obvious
advantage: it is suited for the TeV-scale seesaw mechanism without
turning to tiny Yukawa couplings between the
neutrinos and Higgs doublet: $Y_\nu LH_u\nu^c$. This property is
found to be capable of mitigating the great stress in the Minimal
SSM (MSSM) which, to have the relatively heavy SM-like Higgs boson
mass, incurs a rather serious fine-tuning from generating both the
weak scale and  LSP neutralino dark matter
phenomenology~\cite{Kang:2012sy,Barger}.

To demonstrate the consequence of this property, we consider that
the new (single family of) Yukawa coupling develops an IR-fixed
point, which predicts $Y_\nu\simeq0.75$. This new large Yukawa
coupling involving $H_u$ at the low energy contributes to 
 the lightest CP-even Higgs boson mass $m_h$ radiatively. 
Using the effective potential
method~\cite{Haber:1993an,Carena:1995wu}, we first analytically
calculate such corrections in some simplified cases and then employ
the full numerical analyses. Enhancement up to 4 GeV can be obtained
in the case of a large sneutrino mixing. This helps to alleviate the
tension between a relatively heavy SM-like Higgs boson and the
weak-scale naturalness.

The sneutrino LSP in the SSMs with low-scale seesaw mechanism may
be a good alternative of the neutralino LSP
DM~\cite{sneutrino:early,Dirac:real}, especially after the discovery
of the SM-like Higgs boson and null results from the DM detection
experiments like XENON100~\cite{XENON100}. Specified to the low scale
supersymmetric ISS, the sneutrino LSP is expected to be complex.
This restricts the sneutrino DM into two possibilities: (I)
Essentially it belongs to the Higgs-portal complex DM, and its mass
has to be around $m_h/2$ or above the $W$ boson mass $m_W$; (II) It is
 a coannihilating DM, for example, the sneutrino LSP and 
Higgsino coannihilation. This allows a rather
weak coupling between sneutrino DM and visible particles, so it is
hard to be detected. Given an extra family of sneutrinos, both
scenarios naturally work when we attempt to suppress the left-handed
sneutrino DM component, confronting with enhancing $m_h$.

This paper is organized as follows.  In Section II, we briefly
introduce the model and then calculate the radiative correction to
Higgs boson mass from the neutrino 
 Yukawa coupling at the IR-fixed point. In
Section III, we discuss the complex sneutrino dark matter
phenomenology and investigate how they are consistent with the
requirement of enhancing Higgs boson mass. The Section IV includes
 discussions and conclusion.

\section{ The Lightest CP-Even Higgs boson mass}

The models equipped with a low scale seesaw mechanism receive special
attentions, by virtue of its potential to be tested within our near
future experiments. Specified to the SSMs, after the discovery of a
relatively heavy SM-like Higgs boson, models capable of enhancing
the Higgs boson mass $m_h$ gain further theoretical preference, as
stressed in the introduction. In type-I and III seesaw mechanisms,
where the small neutrino mass $m_\nu\sim Y_\nu^2 v^2/M_R$ with $M_R$
the seesaw scale, the enhancement is impossible~\cite{Hirsch:2012ti}
because a low scale $M_R$ is at the price of a negligibly small
$Y_\nu$~\footnote{In Ref.~\cite{Cao:04}, it was pointed out that  
in type-I seesaw the correction to Higgs boson mass can be significant. 
For a high scale $M_R$ 
(thus allowing $Y_\nu\sim 1$) but at the same time a very large soft mass squared 
for the right-handed sneutrino, unfortunately, the correction is negative.}.
 By contrast, in the type-II and inverse seesaw mechanisms
the smallness of $m_\nu$ has other origins, and then the Higgs
doublets are allowed to have large neutrino Yukawa couplings. For
instance, the supersymmetric type-II seesaw mechanism can even
enhance $m_h$ at the tree level~\cite{Kang:2013wm} (Actually, it can
simultaneously enhance the di-photon rate~\cite{Kang:2013wm}.).
However, such models are difficult to be embedded into a pertubative
Grand Unified Theory (GUT) picture. The inverse seesaw models, where
only singlets are involved, are potential to lift the Higgs boson
mass (at one-loop level) without violating GUT.

In this Section we will first briefly review the minimal
supersymmetric ISS mechanism, and then discuss one
of the new Yukawa couplings with the IR-fixed point behavior and
 its implication to the correction on the SM-like Higgs boson mass.

\subsection{The ISS Model with an IR-Fixed Point}

The minimal supersymmetric model with ISS is the MSSM extended by
two extra singlets $\nu^c$ and $N$ (a single family for the time
being), which carry lepton numbers $-1$ and 1 and are dubbed as
right-handed neutrino (RHN) and Dirac partner RHN (DRHN),
respectively. The superpotential, asides from the ordinary MSSM
terms, can be written as a sum of terms respecting the lepton number
and a term violating it explicitly but slightly:
\begin{align}
W_{ISS} =\L Y_{\nu}{\nu}^c L {H}_u + M_R {\nu}^c N\R + \frac{1}{2}
\mu_N N^2.
\end{align}
The model contains one dimensionless parameter $Y_\nu$, and two
dimension-one mass parameter $M_R$ and $\mu_N$. The $\mu_N$ term softly breaks the lepton
number by two units and largely accounts for the smallness of
tiny neutrino mass ~\footnote{Based on the
next-to-MSSM, the Dirac mass $M_R$ can be dynamically
generated~\cite{Kang:2011wb,Gogoladze:2012jp}. In particular,
Ref.~\cite{Kang:2011wb} observed that, in the presence of a singlet
with a TeV-scale VEV, the small $\mu_N-$term can be related to a
dimension-five operator suppressed by the Planck scale.}.

The model naturally gives rise to a low scale seesaw mechanism
without turning to small parameters except for the massive parameter
$\mu_N$. In the basis ($\nu$, $\nu^c, N$) the neutrino mass matrix
is given by
\begin{align}
\mathcal{M}_F =  \left(     \begin{array}{ccc}
                                    0 & Y_{\nu}v_u & 0  \\
                                     Y_{\nu}v_u  & 0 & M_R  \\
                                   0 & M_R & \mu_N
\end{array}   \right)~,~
\end{align}
where $v_u$ is the Vacuum Expectation Value (VEV) of Higgs field $H_u$,
{\it i.e.}, $\langle H_u \rangle$ = $v_u$. This matrix leads to the following
combination as the light Majorana neutrino
\begin{align}\label{}
\nu_{1}\approx \sin\theta_1 \nu_L-\cos\theta_1 \nu^c,
\end{align}
where the mixing angle is approximated to be $\sin\theta_1\approx M_R/\sqrt{m_D^2+M_R^2}$
with $m_D \equiv Y_{\nu}v_u$ the Dirac mass for $\nu_L$ and $\nu^c$.
The neutrino mass assumes a form of
\begin{align}
m_{\nu} \simeq \frac{m_D^2}{M_R^2}\mu_N~.
\end{align}
Notably, if $\mu_N$, for some reason, can be arbitrarily small,
then the sub-eV neutrino mass scale can be obtained without turning
to the large suppression from extremely small $Y_\nu$ or (and) large
$M_R$. This merit of the ISS model is the basic observation of our
article. But note that owing to the non-unitary constraint, $M_R$
should be several times larger than $m_D$. From
Ref.~\cite{Valle:2004,unitarity} we set a rough bound
\begin{align}\label{}
M_R\gtrsim 10m_D~,~\,
\end{align}
so as to make $\theta_1\simeq\pi/2$, i.e., the neutrino is dominated
by left-handed neutrino. Finally, $\nu^c$ and $N$ form a Dirac
fermion with mass approximately given by
\begin{align} \label{fermionsquaremass}
M_{R} + {m_D^2}/{2M_R}~.
\end{align}

\begin{figure}
\begin{center}
\includegraphics[scale=1]{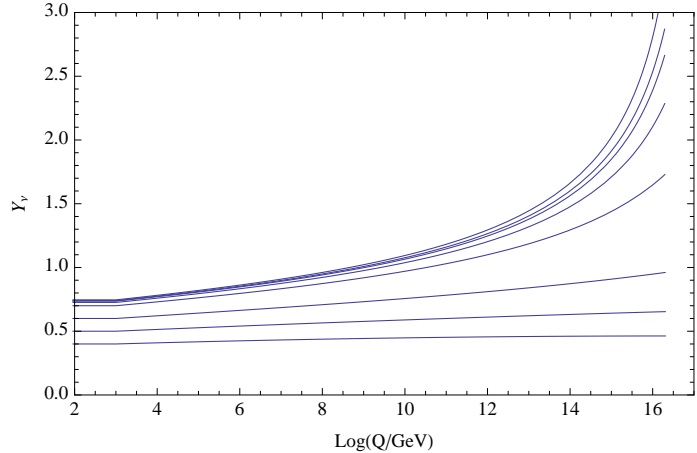}
\caption{\label{yukawa} Renormalization group equation running trajectory for
$Y_{\nu}$ in the inverse seesaw model, showing that the Yukawa coupling has an
infrared-stable quasi-fixed point for large $Y_{\nu}$. Here, $m_t = 173.1$ and $\tan\beta
= 10$ are chosen.   }
\end{center}
\end{figure}

For a reason discussed later, we are interested in how large
$Y_\nu$ is allowed by pertubativity up to the GUT scale.
Interestingly, we find that there is an IR-fixed point structure
predicting $Y_\nu\lesssim0.75$.
We present the Yukawa running in this model and focus on the new
Yukawa term which is illustrated in Fig.~\ref{yukawa}. For
numerical calculation, we choose the SUSY-breaking scale $m_{S} = 800 $
GeV, new sterile neutrino scale $M_R = 1000$ GeV, and we simply
consider one generation new sterile neutrinos. The new Yukawa
coupling has infrared quasi-fixed point behavior, which restrict how
large it can be at the TeV-scale while maintaining consistent with
 perturbative unification. Here for fixed-point trajectory we
adopt the same definition like in \cite{yukawafix} which request the
new Yukawa couplings is less than or equal to 3.

\subsection{Stau-sterile neutrino contribution to the Higgs boson mass}

It is well known that  at tree level the
SM-like Higgs boson mass $m_h$ is predicted to be lighter than $M_Z$
in the MSSM. This necessitates an
significant radiative correction from the stop-top sector to lift
the Higgs boson mass near 125 GeV, which is discovered by the recent CMS and ATLAS
Collaborations~\cite{LHC:Higgs}. However, such large correction
typically requires stops at the TeV scale and then renders the MSSM
highly fine-tuned~\cite{Kang:2012sy}. Therefore, how to lift the
Higgs boson mass at the less price of fine-tuning is a
very interesting question.

The supersymmetric ISS model with the IR-fixed point just provides a
new source of enhancement via the stau-sterile neutrino correction.
We calculate the corrections following the effective potential
method~\cite{Haber:1993an,Carena:1995wu} from the
Colenman-Weinberg potential~\cite{CW} of the Higgs field via a general
formula
\begin{align}
\Delta V(H_u) =& 2\sum_{i}\left[ F(M_{S_i}^2) - F(M_{F_i}^2)
\right],\cr F(M^2) =&\f{ M^4}{64\pi^2}\left[ \ln(M^2/Q^2)
-3/2\right],
\end{align}
with $Q$ the renormalization scale. Index $i$ runs over all the
scalar and fermion couplings to Higgs doublets. Restricted to the
neutrino sector, the fermion spectrum contains a light Majorana
neutrino, whose contribution to the effective potential is
proportional to $(\mu_N/M_R)^4$ and thus can be safely dropped.
While the Dirac sterile neutrinos have a large mass given by
Eq.~(\ref{fermionsquaremass}). We postpone the discussion on the
scalar spectrum to the next paragraph. Now, in the decoupling
region, {\it i.e.}, $m_{A^0}^2 \gg m_{h}^2$,
the effective potential gives the following correction to $m_{h}^2$ 
 \begin{align} \label{mhdelta}
 \Delta m_{h}^2 =\left \{ \frac{\sin^2 \beta}{2}\left[ \frac{\partial^2}{\partial v_{u}^2} -
 \frac{1}{v_u}\frac{\partial}{\partial v_u} \right]  +
 \frac{\cos^2 \beta}{2}\left[\frac{\partial^2}{\partial v_{d}^2} - \frac{1}{v_d}\frac{\partial}{\partial
 v_d} \right]
 + \sin\beta\cos\beta\frac{\partial^2}{\partial v_u \partial v_d} \right\} \Delta V.
\end{align}

We now turn our attention to the scalar spectrum. First,
 the supersymmetric ISS model introduces new soft terms
{\small\begin{align} - \mathcal{L}_{soft} =-\mathcal{L}_{soft}^{\rm
MSSM} + m_{\wt\nu^c}^2| \tilde{\nu}^c |^2 + m_{\wt N}^2 | \tilde{N}
|^2 +\L Y_{\nu}A_{Y_{\nu}}\tilde{L}\tilde{\nu}^c H_u +
B_{M_R}\tilde{\nu}^c\tilde{N} +
\frac{1}{2}B_{\mu_{N}}\tilde{N}^2+c.c.\R~,
\end{align}}where $\mathcal{L}_{soft}^{\rm MSSM}$ is the MSSM
SUSY breaking soft term. $B_{M_R}\sim m_{3/2} M_R$ with $m_{3/2}$
the gravitino mass measuring the scale of soft breaking parameters.
It is noticed that in our scenario the lepton number violating soft
breaking parameter $B_{\mu_N}\sim m_{3/2}\mu_N$ is irrelevantly
small (we will come back to this point later), and consequently 
three sneutrinos are highly complex scalars. Now, in the basis
$(\wt\nu_L,(\wt \nu^c)^*,\wt N)$ the sneutrino mass squared matrix
takes a form of
\begin{align} \label{scalarsquaremass}
\mathcal{M}_{S}^2 = \left(
                         \begin{array}{ccc}
                                           m_D^2+m_{\wt L}^2 + \frac{1}{2}m_{Z}^2 \cos 2\beta & m_D (A_{Y_{\nu}} - \mu  \cot \beta)  & m_DM_R   \\
                                      &  m_D^2+M_R^2+ m_{\wt\nu^c}^2  & B_{M_R}  \\
                                      &  &   M_R^2+m_{\wt N}^2
                         \end{array}  \right).
\end{align}
Because the analytical eigenvalues are extremely lengthy, in the ensuing discussion
we will consider some solvable
limits to demonstrate which parameters can lift the Higgs
boson mass.

There are two mixing terms which depend on $H_u$ and thus contribute
to $m_h^2$ in terms of Eq.~(\ref{mhdelta}), one is the soft
trilinear term $X_\nu  = A_{Y_{\nu}} v_u - \mu \cot \beta$,
while the other one comes from the F-term of $\nu^c$, {\it i.e.}, $Y_\nu
v_u M_R=m_DM_R$ (In light of the effective potential method, any
origin of correction to $m_h$ can be traced back to the matrix
entries depending on $m_D$). Turning off the mixing $A$-term will lead
to a solvable matrix. As a warm up, we first consider the case by
turning off the $A_\nu$ and $B_{M_R}$ mixing terms, setting the soft
SUSY-breaking mass squares equal to $m_S^2$ and neglecting the small
electroweak D-term contribution. Then in the mass eigenstates the
three sneutrino mass squares are given by
\begin{align} \label{bosonmass}
 m_{S}^2 ,~~~ m_{D}^2 + M_{R}^2 +m_{S}^2, ~~~m_{D}^2+
 M_R^2+m_{S}^2.
\end{align}
The eigenstate corresponding to $m_S^2$, which is independent on
$H_u$, does not contribute to the effective potential. But it is
merely
a result of the approximations which we have taken. 
 Using Eq.~(\ref{mhdelta}) we get
\begin{align}\label{nomixing}
\Delta m_{h}^2 = \frac{1}{4 \pi^2}\sin^2\beta Y_{\nu}^2
m_{D}^2\ln\frac{M_{R}^2+m_{S}^2}{M_{R}^2}.
\end{align}
It is similar to the non-mixing stop case but quantitatively less
important due to a color factor and the smaller Yukawa
coupling which leads to a suppression $(Y_\nu/h_t)^4$. Even worse is
that, from the neutrino side we have $M_R\sim 1$ TeV and
consequently here the logarithmic enhancement is rather limited if
$m_S$ is constrained around the TeV scale by naturalness. Thus, the
correction from sneutrino is less than 10 percents of that from the stop case,
see the left panel of Fig.~\ref{comparison}.

 \begin{figure}
 \centering
 \includegraphics[scale=1]{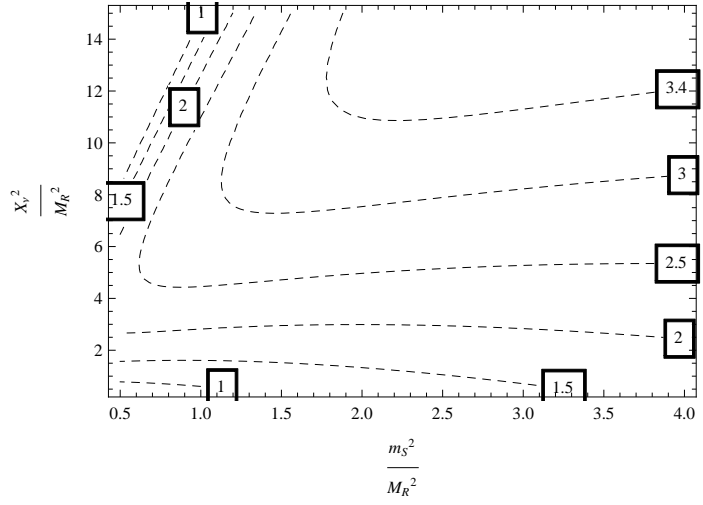}
 \caption{\label{contourplot} $\frac{X_\nu ^2}{M_R^2}$ versus $\frac{m_S^2}{M_R^2}$ 
 for the contour of $\Delta m_h$, which is defined as 
 $\sqrt{(123{\rm\,GeV})^2 + \Delta m_h^2} - 123\rm\,GeV$ throughout this work.}
 \end{figure}
Next we take into account the soft trilinear term $X_\nu$, which
would bring much difference in lifting $m_h$. If other
approximations are the same as the non-mixing case, we are still
able to find an analytical expression for the correction, in spite
of somewhat complication
 \begin{align}\label{total}
\Delta m_{h}^2 =  \frac{1}{4\pi^2} \sin^2\beta Y_{\nu}^2 m_{D}^2
 &\left( \ln\frac{ m_{S}^2 +M_{R}^2}{M_{R}^2} + \frac{X_\nu ^4+2X_\nu ^2M_R^2}{M_R^4}
   \right.& \nonumber \\
   &\left.     - \frac{2X_\nu ^4m_S^2+X_\nu ^4M_R^2+
    4X_\nu ^2m_S^2M_R^2}{2M_R^6} \ln\frac{m_S^2+M_R^2}{m_S^2}
\right).
 \end{align}
We would like to comment on the origins of various terms in
the above equation~\footnote{We note that
 our approximation expression  is different from that
 in Ref.~\cite{Gogoladze:2012jp} which is consistent with that in Ref.~\cite{yukawafix}.
 However, from our understanding, the particle contents and interactions in
 their inverse seesaw models are not the same as the extended MSSM with
 extra vector-like particles.}.
The first logarithmic term exactly reproduces the result given in
Eq.~(\ref{nomixing}), extracting the correction to the Higgs quartic
coupling $\ld_h$, which is encoded in the 
Renormalization Group Equation (RGE) running from the
supersymmetry breaking scale defined by the heavy sneutrino mass
scale to the Dirac sterile neutrino mass scale $M_R$.  While the
second term stands for the mixing effect after integrating out the
sneutrinos, included as a shift to the boundary $\ld_h$ (In the
explicit Feymann diagram calculations, it can be obtained from the
triangle and box diagrams.). The last term has an obscure dependence
on the logarithmic $\log ((m_S^2+M_R^2)/m_S^2)$, stemming from the
different mass scales of $\wt \nu_L$ and $\wt \nu^c$ as shown in
Eq.~(\ref{scalarsquaremass}). Such a hybrid of the mixing and
logarithmic terms is absent in the stop system, and noticeably it is
negative.


We now quantitatively analyze the corrections given in
Eq.~(\ref{total}). $m_D$ has been fixed by virtue of the IR-fixed of
$Y_\nu$, and then it is not difficult to find that $\Delta m_h^2$
actually depends on two dimensionless parameters,
$x_s={m_S^2}/{M_R^2}$ and $x_a={X_\nu^2}/{M_R^2}$. Denoting the
function in the brackets as $f(x_s,x_a)$, it consists of three parts:
the first and second parts are positive while the third part is
negative. Hence the maximal mixing scenario here is more subtle than
the stop case. It is illustrative to consider two limits of $x_s$:
(1) If $x_s\gg1$, i.e., the large SUSY breaking  soft mass terms, we
have a simple expression for $f(x_s,x_a)\ra \log x_s-x_a^2/2x_s$.
Thereby the mixing effect is negative but suppressed by large $x_s$.
(2) Oppositely, if $x_s\ll 1$ we have $f(x_s,x_a)\ra x_a^2(1+\log
\sqrt{x_s})+2x_a$. Thus, a moderate $x_s$ is needed to maximize the
corrections. Explicitly, we present a contour plot of $\Delta m_h$ in
Fig.~\ref{contourplot}.
 \begin{figure}
 \begin{center}
 \includegraphics[scale=0.67]{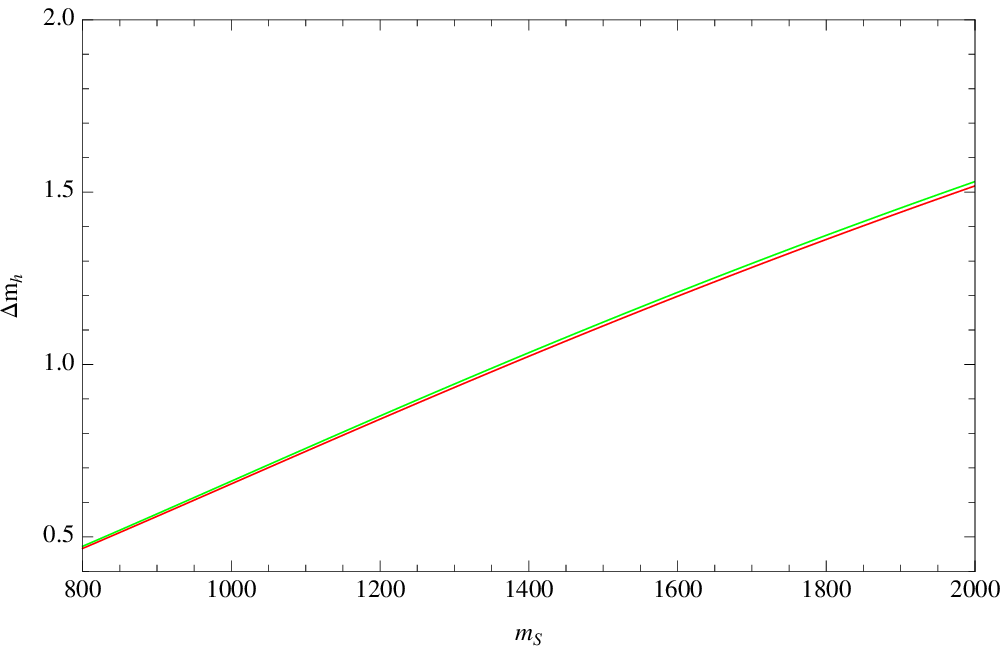}
 \includegraphics[scale=0.75]{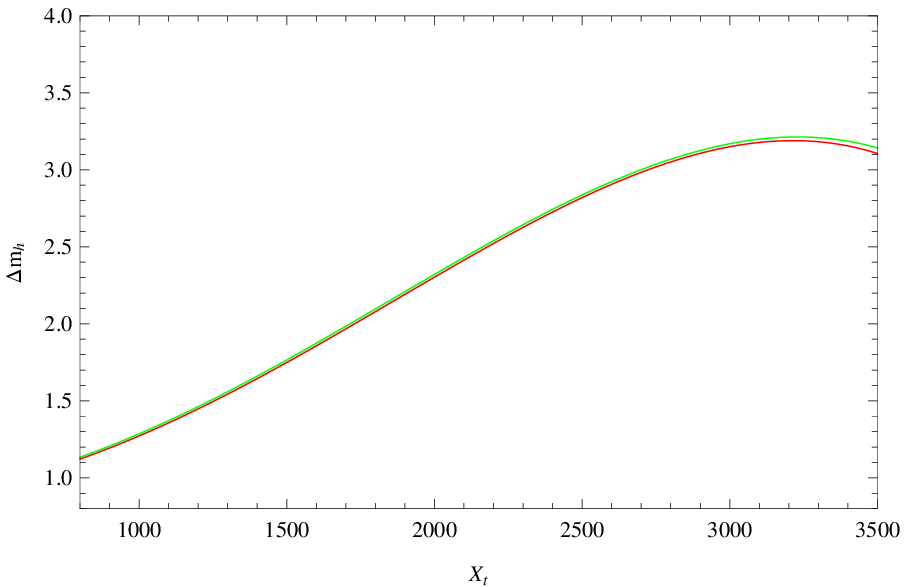}
 \caption{\label{comparison} Left panel: $\Delta m_h$ versus $m_S$ with vanishing
soft trilinear term. Right panel: $\Delta m_h$ versus $X_\nu $ for
 $M_{R} = 1000$ GeV, $m_S = 800$ GeV, and $\tan \beta = 15$.
The red lines correspond to the full numerical results and green
lines the approximated results.}
 \end{center}
 \end{figure}

\begin{figure}
\begin{center}
\includegraphics[scale=0.8]{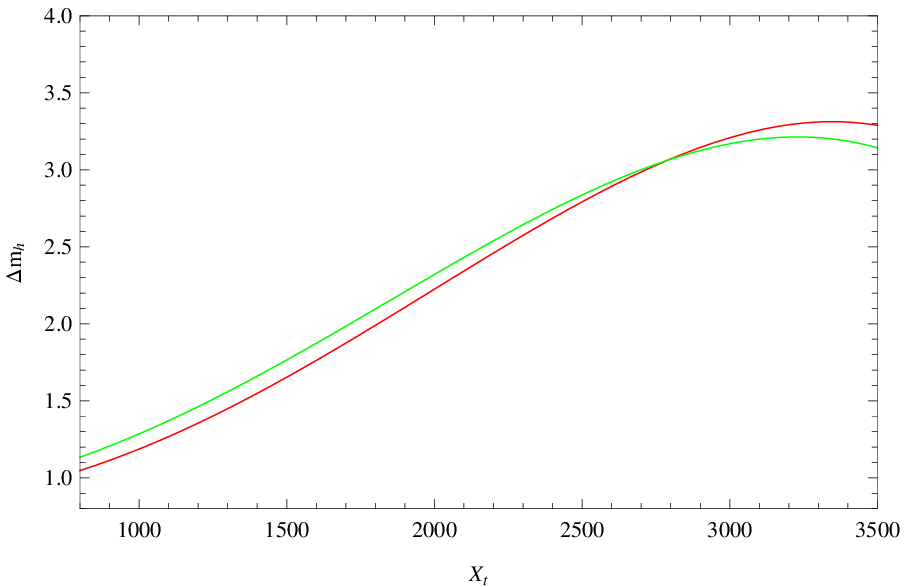}
\includegraphics[scale=0.8]{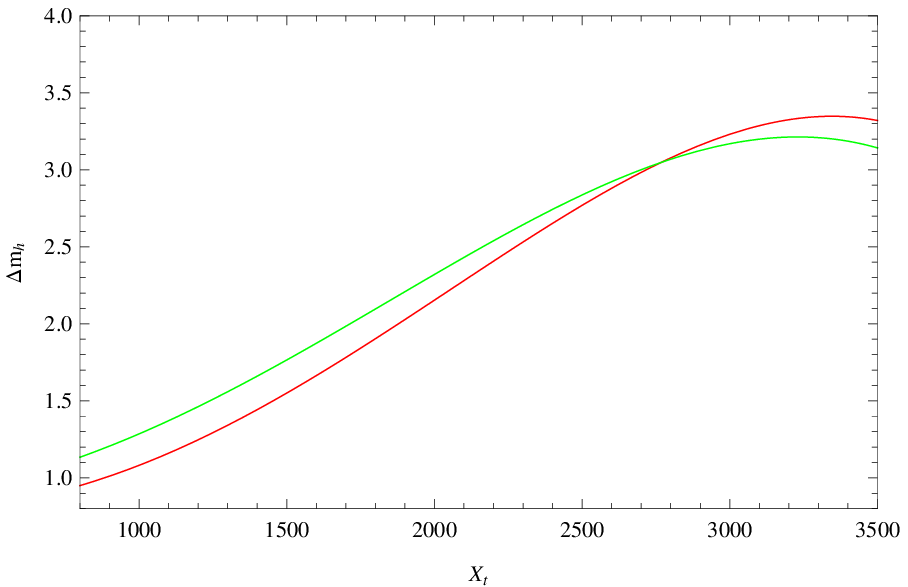}
\caption{\label{fullcomparison} Left: $\Delta m_h$ versus $m_S$ for
the full numerical results with $B_{M_R} =0.5m_S^2$; Right: $\Delta
m_h$ versus $m_S$ for the full numerical results with $B_{M_R} =
m_S^2$. We have fixed $M_{R} = 1000$ GeV, $m_S = 1200$ GeV, and $\tan
\beta = 15$.}
\end{center}
\end{figure}

We examine the difference between the approximately analytical and
full numerical treatments. Fig.~\ref{comparison} shows that they
give rise to almost the identical results, in the trivial case of
non-mixing (left panel) and the case with $A_\nu$ mixing only (right
panel). This indicates that the expression in Eq.~(\ref{total}) works
well for universal soft masses. In Fig.~\ref{comparison}, for
simplicity we set $B_{M_R} \sim$ 0. However, a non-zero $B_{M_R}$ may
lead to an appreciable change. Although it is still possible to
develop an analytical expression for $\Delta m_h$, it is too
lengthy to convey any useful information. Therefore, we only display
the change numerically in Fig.~\ref{fullcomparison}, with color
codes of lines as before. It is clearly seen that, as
$|B_{M_R}|^{1/2}$ approaches the sneutrino mass scale, the
discrepancy between the full result and the approximation
Eq.~(\ref{total}) becomes rather significant.

\section{A Complex Sneutrino LSP in the ISS-MSSM}

The presence of a DM candidate, {\it i.e.}, the neutral LSP, is one of the
major attractions of the SSMs. The lightest neutralino receives the most
intensive attention. However, the heaviness of the soft spectrum,
owing to a relatively heavy Higgs boson and null sparticle searches,
along with the stringent bounds from direct detection experiments
such as XENON100~\cite{XENON100}~\footnote{After the completion of this 
work, the LUX Collaboration announced their new results, which gave the
even more stringent constraint, one magnitude of order stronger than 
the XENON100 experiment. 
But this does not qualitatively affect our discussions, so we only 
mention XENON100 in the text and show LUX data in figures.}, 
now threaten the viable
neutralino DM. Another neutral LSP candidate, the lightest
sneutrino, has been investigated by many authors in many
contexts~\cite{sneutrino:early,sneutrino:Z,Arina,Dirac:real,Mondal:1306}, and now
may show some advantages.

Roughly speaking, the sneutrino LSPs can be classified into the following three types
\begin{description}
  \item[Complex sneutrino] As in the MSSM, the left-handed sneutrino
  $\wt\nu_L$, just the original proposal~\cite{sneutrino:early}, is
  a complex scalar due to the conservation of global lepton number
  $U(1)_L$. But the $Z-$boson mediated DM-nucleon spin-independent (SI)
  scattering has a very large cross section
  $\sigma_{\rm SI}$~\cite{sneutrino:Z}, which excludes the
  sneutrino LSP in the MSSM. Beyond it, a complex sneutrino LSP
  dominated by the SM singlets may be realized in the models which
  conserve the lepton number with a high
  degree~\cite{Dirac:sneutrino,Kang:2011wb}.
   \item[Real sneutrino] Taking into account for generating tiny
   neutrino masses via the seesaw mechanisms,  $U(1)_L$
  should be broken and then induces the mass splitting between the CP-odd and even
   components of complex sneutrino. If splitting
   is large, we actually have a real sneutrino LSP~\cite{Arina}.
   Consequently, $\sigma_{\rm SI}$ from $Z-$boson exchange will be zero.
    \item[Pseudo-complex sneutrino] When the splitting is small,
    which is the usual case because of suppression from small
    neutrino mass, we will get a pseudo-complex
    sneutrino~\cite{Arina,inels:sneutrino,An:2011uq}.
    In this case, the DM-nucleon scattering mediated by $Z-$boson
    becomes inelastic and may be kinematically forbidden.
\end{description}
In the low-scale ISS-MSSM, $U(1)_L$ is broken by $\mu_N N^2/2$ and
the corresponding soft bilinear term $B_{\mu_N}\wt N^2/2=m_{\rm
SUSY} \mu_N\wt N^2/2$. Provided that there is no peculiar
SUSY-breaking mediation such as in Ref.~\cite{Dirac:sneutrino}, the
mass splitting between the CP-even and CP-odd parts of $\wt
N_1$, due to the soft bilinear term, is suppressed by the neutrino
mass
\begin{align} \label{}
\delta m\sim \f{m_{\rm SUSY} \mu_N}{2m_{\wt N_1}}\sim \L\f{m_{\rm
SUSY} }{2m_{\wt N_1}}\R\L\f{M_R^2}{m_D^2}\R m_\nu,
\end{align}
which is around 10 eV. Thereby, in this model the sneutrino LSP is
expected to be complex~\cite{Kang:2011wb}. Moreover, it is thermal
since it is free of extreme suppression from Yukawa couplings with
the visible sector. But note that enhancing $U(1)_L-$violation by a
abnormally large $B_{\mu_N}\sim {\cal O\rm\,(GeV}^{2})$ can lead to
an inelastic sneutrino DM, which actually is the case in most
references~\cite{Arina:2008bb,An:2011uq,DeRomeri:2012qd}. In this
article, we insist on a normal SUSY breaking mediation mechanism to
account for soft terms, and then we have a (highly) complex
sneutrino LSP. Our purpose is to explore the viable sneutrino DM
scenarios consistent with the enhancing Higgs boson mass.

\subsection{General Simplifications due to Complexity}\label{general}

We investigate what kind of mixture can lead to a good sneutrino DM,
which has correct relic density, is allowed by the XENON100
experiment and does not
incur tremendous fine-tuning. In general, for the moment we
do not restrict discussions to the setup made in the previous
Section. From the first glance, the sneutrino LSP is complicated because of
three complex sneutrino system, but the complexity of the sneutrino LSP
greatly simplifies the discussions.

Above all, we have to suppress the left-handed sneutrino fraction.
The lightest sneutrino is a superposition of three sneutrinos
\begin{align} \label{comp}
\wt N_1=C_{\wt \nu_L}\wt \nu_L+C_{\wt \nu^c}(\wt \nu^c)^*+ C_{\wt N}
{\wt {N}},
\end{align}
where the singlet fraction must dominate to suppress the $Z-$mediated
DM-nucleon scattering. It is justified to make an estimation on the
doublet fraction
\begin{align} \label{C:esti}
C_{\wt \nu_{L}}\simeq
\f{m_{D}A_{\nu}}{m_{\wt\nu_L}^2}\sin\theta_{23}+
\f{m_{D}M_{R}}{m_{\wt\nu_L}^2}\cos\theta_{23},
\end{align}
which is quite precise in the case of large mass splitting between
$\wt\nu_L$ and other two sneutrinos. $\theta_{23}\in (-\pi/2,0)$ is
the mixing angle of $\wt \nu^c$ and $\wt N$ from the 23-submatrix of
Eq.~(\ref{scalarsquaremass})
\begin{align} \label{}
\tan\theta_{23}\approx P_{23}- \sqrt{ P_{23}^2+1 },\quad
P_{23}\equiv \L{m_{D}^2+m_{\wt \nu^{c}}^2-m_{\wt
N}^2}\R/{2B_{M_{R}}}.
\end{align}
In Eq.~(\ref{comp}) $C_{\wt \nu^c}\approx -\sin\theta_{23}$ and
$C_{\wt N}\approx -\cos\theta_{23}$.
 The DM-proton SI
scattering cross section is $\sigma_p={\mu_p^2}
f_p^2/\pi$~\cite{SUSYDM}, with $\mu_p$ the DM-proton reduced mass.
For a DM with mass around 100 GeV, XENON100 imposes the upper bound
$f_p\lesssim 0.3\times10^{-8}$ GeV$^{-2}$ and in turn
\begin{align} \label{}
f_p=C_{\wt \nu_L}^2
g_2^2/m_Z^2\lesssim0.3\times10^{-8}{\rm\,GeV^{-2}}\Rightarrow C_{\wt
\nu_L}\lesssim0.01.
\end{align}
A natural suppression needs a multi-TeV $\wt\nu_L$ for
$Y_{\nu}\sim{\cal O}(0.1)$. By contrast, a much lighter $\wt\nu_L$
(but still heavier than $\wt N$ and $\wt \nu^c$) is allowed given a
sufficiently small $Y_{\nu}$.

With such a small $C_{\wt \nu_L}$, the dynamics of the
sneutrino LSP largely reduces to that of a Higgs-portal complex scalar DM.
Concretely, $\wt N_1$ annihilates into SM particles through four
ways: (1) Contact interactions with $H_u$ from $|F_{\nu_L}|^2$; (2)
$h$ propagating in the $s-$channel; (3) Higgsinos/sneutrinos
propagating in the $t-$channel; (4) Gauge interactions inheriting
from $\wt\nu_L$. Decoupling $\wt\nu_L$ means only case (1) left, giving
rise to a Higgs-portal complex scalar DM
\begin{align} \label{contact}
|F_{\nu_L}|^2\ra \ld_H|\wt N_1|^2|H_u|^2,\quad \ld_H\equiv
(\sin\theta_{23}Y_{\nu})^2.
\end{align}
But cases (2) and (3) may cause deviations from an exact Higgs-port DM. In
the first,  the Higgsino-mediated processes might be important for
light DM well below $m_W$, since their cross sections are $\sim
{\cal O} (\ld_H^2 m_{\wt N_1}^2/32\pi M_{\rm Higgsino}^4)$, with further
velocity/helicity suppressions~\cite{Kang:2011wb}.
But owing to the bound indicated in Eq.~(\ref{H:bound}), we find
that it is far from enough to give the correct relic density and
thus can be ignored. Next, if $C_{\wt \nu_L}$ is not extremely
suppressed, processes involving $\wt\nu_L$ contributions to DM cross
sections at higher order of $C_{\wt \nu_L}$ can be enhanced
by large massive couplings. This is seen from the following terms,
which are absent in the ordinary Higgs-portal DM models,
\begin{align} \label{devi}
-{\cal L}_{trilinear}=&|F_{\nu^c}|^2+\L Y_\nu A_\nu \wt LH_u\wt
\nu^c + c.c.\R \cr \ra & -\f{\L C_{\wt \nu_L} m_{\wt
\nu_L}\R^2}{\sqrt{2}v_u}h|\wt N_1|^2-C_{\wt \nu_L}\f{m_{\wt
\nu_L}^2}{\sqrt{2}v_u}h\wt\nu_L\wt N_1^*+c.c.
\end{align}
The presence of extra $C_{\wt \nu_L}$ in the first and second terms
are traced back to the fact that these trilinear couplings are also
sources of the mixing terms between $\wt\nu_L$ and singlets, see
Eq.~(\ref{scalarsquaremass}).

We proceed to argue that terms in Eq.~(\ref{devi}) can make no
difference. The first term of Eq.~(\ref{devi}) affects processes
mediated by $h$. With it, the massive coupling constants of $h|\wt
N_1|^2$, $\mu_h$, takes a form of
\begin{align} \label{H:bound}
\mu_h=\sqrt{2}v_u \ld_H\left[1+ \L C_{\wt \nu_L}m_{\wt
\nu_L}/\sqrt{2\ld_H}v_u \R^2\right].
\end{align}
The second term stands for the deviation from the exact
Higgs-portal. In terms of Eq.~(\ref{C:esti}), its order of magnitude
is estimated to be at least $\sim {\cal O}$$(A_\nu/\sqrt{2}m_{\wt\nu_L})^2$, 
which is reliable barring the large
cancellation in Eq.~(\ref{C:esti}). So, for a relatively large
$A_\nu$ but small $m_{\wt \nu_L}$, which means an appreciable
$C_{\wt \nu_L}$, the trilinear soft term dominates $\mu_h$.
$\sigma_{\rm SI}$ from $h$ implies the bounds
\begin{align} \label{H:bound}
(C_{\wt \nu_L}m_{\wt\nu_L})^2/\sqrt{2}v_u m_{\rm DM},\, \sqrt{2}v_u
\ld_H/m_{\rm DM}\lesssim \mu_h/m_{\rm DM}\lesssim 0.12
\times(\sigma_p^{\rm up}/10^{-9}\rm\,pb)^{1/2}.
\end{align}
But the dynamics of the Higgs-portal DM with and without deformation (by
trilinear soft terms) are the same, provided that annihilation into
a pair of $h$ via the contact interaction in Eq.~(\ref{contact}) is
irrelevant. Explicitly, it possesses $\sigma^c_{hh}
v=\L\sqrt{2}\ld_H v_u/m_{\rm DM}\R^2/128\pi v_u^2\lesssim 0.5$ pb, a
marginally relevant value. Thus, this deviation does not matter much.
And the second term of Eq.~(\ref{devi}) is not important neither. It
gives rise to the sneutrino-mediated annihilation into a pair of $h$,
with a thermal averaged cross section
\begin{align} \label{}
\sigma_{hh}^t v\simeq \f{1}{32\pi}\L \f{C_{\wt
\nu_L}^2m_{\wt\nu_L}^2}{\sqrt{2}v_u m_{\rm DM}}\R^2 \f{1}{v_u^2}.
\end{align}
Even if the value in the bracket saturates its upper bound shown in
Eq.~(\ref{H:bound}), this cross section merely gives 1 pb. Actually,
it can never be saturated for $m_{\wt \nu_L}\lesssim 1$ TeV and
$m_{\rm DM}>m_h$. In summary, the complex sneutrino DM is
reduced to the Higgs-portal DM.

\subsection{The Sneutrino LSP Confronting with Enhancing Higgs Boson Mass}

With the above analyses, in this subsection we explore the viable
scenarios for sneutrino DM, taking into account the requirement
to lift the Higgs boson mass. But a single family of sneutrino
fails, because of the contradiction between these two aspects. On the one
hand, to significantly lift $m_h$ we need $Y_\nu\sim1$ and thereby
$10\,m_D\lesssim M_R\sim{\cal O}(1)$ TeV. On the other hand, it
renders sneutrinos heavy. Moreover, the mixings between $\wt \nu_L$ and
$\wt\nu^c$/$\wt {N}$ are large by virtue of the large
off-diagonal entries, {\it i.e.}, $m_DX_\nu\wt\nu_L^\dagger \wt \nu^c$ and
$m_DM_R \wt\nu_L^\dagger \wt N$, in the sneutrino mass matrix
Eq.~(\ref{scalarsquaremass}). Therefore, without large fine-tuning,
the single family case can not lift $m_h$ sizably and at the same
time has a sneutrino LSP around the weak scale with a
sufficiently small $\wt \nu_L$ fraction. An extra family of
sneutrino, which has Yukawa coupling $Y_{\nu_2}$ for
definiteness~\footnote{As a matter of fact, to produce the realistic
neutrino masses and mixings, we need extra families as well.}, is thus
introduced.

Now good scenarios can be accommodated. Because $Y_{\nu}$, the third
family Yukawa coupling, has approached the IR-fixed point,
$Y_{\nu_2}$ should be much smaller than it. As a result,
$\wt\nu^c_2$ and $\wt{N}_2$ can be around the weak scale. While
$m_{\wt \nu_{L,2}}^2$ is assumed to be properly heavier and hence
the lightest sneutrino $\wt N_1$ is dominated by $\wt N_2$ and/or
$\wt \nu^c_2$. According to the analysis made in the above
subsection, a sufficient suppression of the doublet component can be
realized via a heavy $m_{\wt \nu_{L,2}}$ for large $Y_{\nu,2}$ or a
small $Y_{\nu,2}$ for light $m_{\wt \nu_{L,2}}$. Both are well
motivated and will be discussed respectively in the following.
Hereafter, the subscript ``2" will be dropped and sneutrinos refer
to the second family unless otherwise specified.

\subsubsection{Higgs-Portal Sneutrino DM Inspired by Natural SUSY}

In the natural SUSY framework, the first and second families of
sfermions are assumed to be much heavier than the third family, says
lying above 3 TeV. Such a pattern may be related to the SM fermion
flavor structure~\cite{Nelson:1997bt}. While sterile neutrinos are
SM singlets and have different flavor structure, and hence their
superpartners are allowed to be light. As a result, we naturally get
a light sneutrino LSP with negligible $\wt\nu_L$ component.

\begin{figure}[htb]
\includegraphics[width=3.3in]{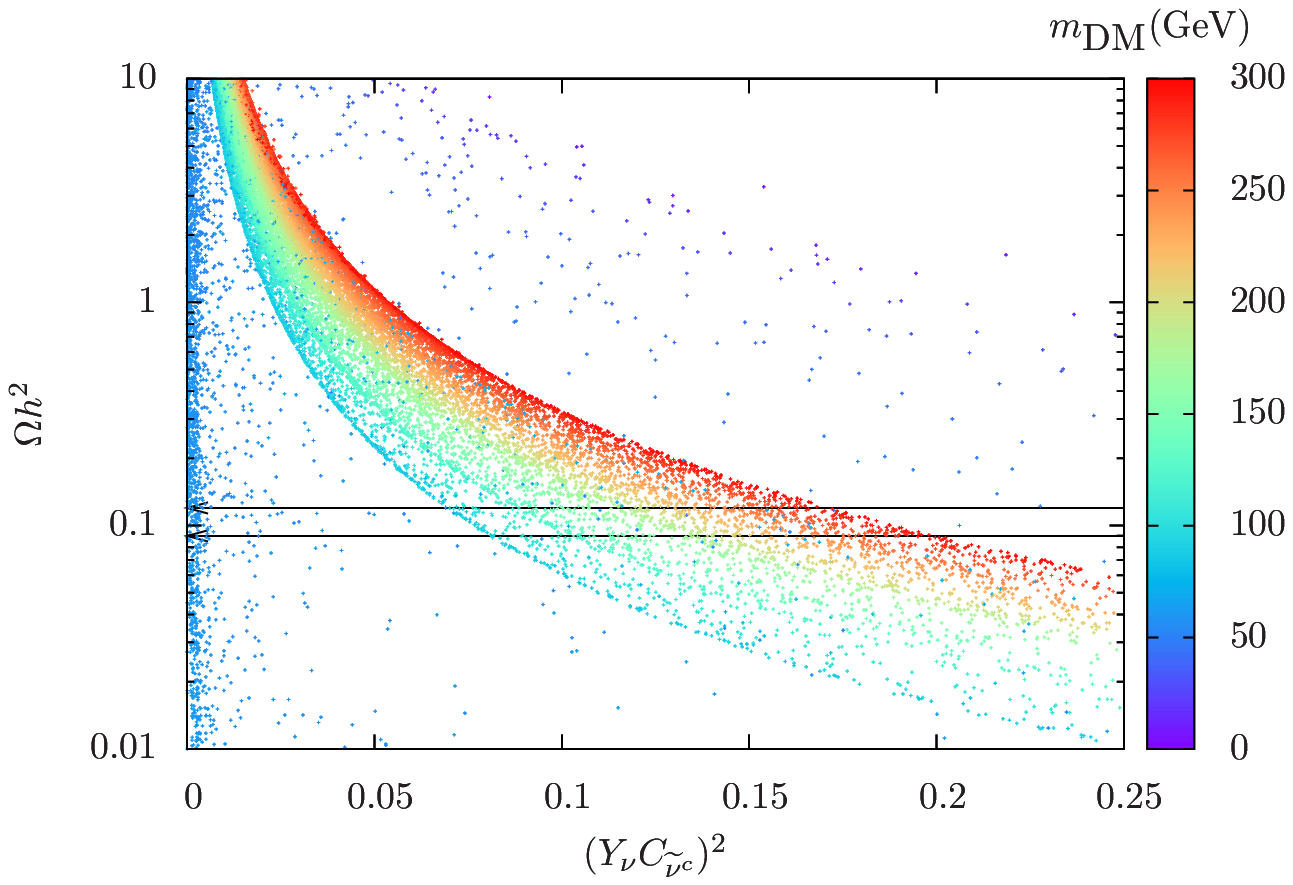}
\includegraphics[width=3.1in]{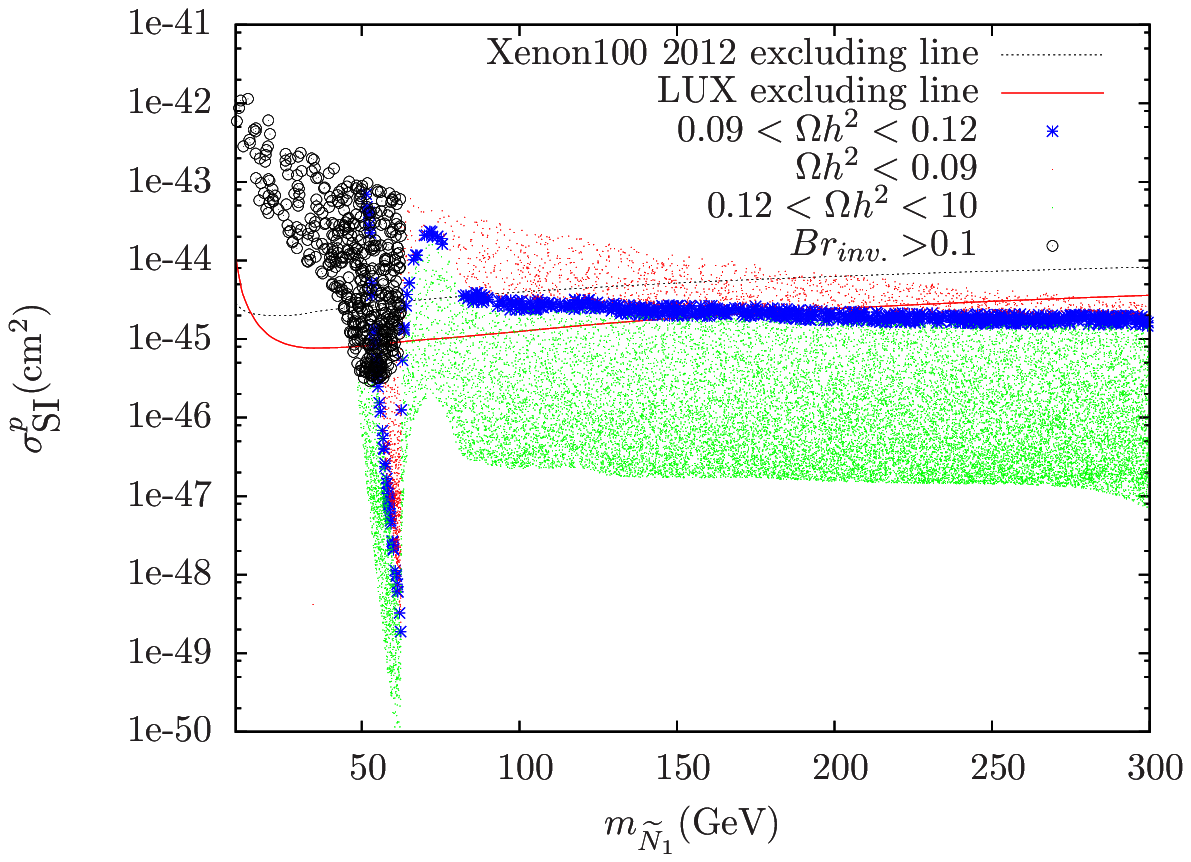}
\caption{Left: The DM relic density versus the effective coupling
constant between the up-type Higgs doublet and sneutrino LSP. Points
in the narrow band between the two black lines have relic density
consistent with the measurement: $0.09<\Omega h^2<0.12$. Right:
The Direct detection exclusion limits on the sneutrino LSP in the
$\sigma_{\rm SI}^p-m_{\rm DM}$
 plane. Assuming the observed DM relic density
can be realized from non-thermal production and dilution mechanism,
we also show the points with smaller and
larger relic densities, labeled as the red and green points,
respectively. Black circles denote the points with
branching ratio of Higgs invisible decay $\geq10\%$.}\label{fig5}
\end{figure}

We are working in the Higgs-portal sneutrino DM, so its correct
relic density, viewing from the stringent XENON100 experimental
constraint, needs to be studied carefully. As mentioned before, the
cross section from DM annihilating into a pair of Higgs bosons via
contact interaction is not large enough. Consequently, annihilations
of the viable sneutrino LSP are completely specified by $h$ in the
$s-$channel. Correct relic density restricts the viable sneutrino LSP 
only to two possibilities (see the left panel of Fig.~\ref{fig5})
\begin{itemize}
\item $m_{\wt N_1}\simeq m_h/2$. When the DM mass lies below the $W$
threshold and closes to the Higgs pole, DM annihilations
will benefit from the Higgs resonance enhancement. Then correct
relic density can be got for a small $\mu_h$. In turn, the XENON100
bound can be evaded, see the right panel of Fig.~\ref{fig5}. The
Higgs invisible decay may impose an even more stringent constraint
for $m_{\rm DM}<m_h/2$. In the right panel of Fig.~\ref{fig5} we
label the points giving invisible Higgs decay to a pair of $\wt N_1$
with branching ratio larger than $10\%$, which, assumed to be the
upper bound, excludes $\wt N_1$ below 55 GeV.
  \item $m_{\wt N_1} \gtrsim m_W$. Bare in mind
that the SM-like Higgs boson decay is always dominated by the $WW$
mode for $m_h\gtrsim  160$ GeV~\cite{Djouadi:2005gi}, thus $\wt N_1$
will dominantly annihilate into $WW$ once it is kinematically
allowed. Increasing DM mass will decrease the DM annihilation cross
section, but new accessible channels such as $ZZ/hh/t\bar t$ can
partially compensate the decrease. As a result, even without
significantly increasing $\mu_h$, the sneutrino LSP with mass
extending above a few times of $m_W$ still can acquire correct relic
density, see the left panel of Fig.~\ref{fig5}. On the other hand, a
heavier scalar DM helps to reduce $\sigma_{\rm SI}$ from $h$, so
the XENON100 experimental constraint is satisfied for the heavier $\wt N_1$, see Fig.~\ref{fig5}.
Actually, from it we see that the latest LUX result has excluded the
sneutrino DM between $65$ and 150 GeV.
\end{itemize}
We modify MicrOMEGAs 3.2~\cite{micrO} by including the ISS-MSSM and
then using it to calculate the sneutrino DM relic density and
$\sigma_{\rm SI}$. Based on the previous semi-analytical analysis,
we scan the following three-dimensional parameter space
\begin{align} \label{}
&A_\nu=200{\rm\,GeV},\quad Y_{\nu}\in [0.05,\,0.3],  \cr &
m^2_{\widetilde{N}}\in [-M_R^2,\,M_R^2] ,\quad
m^2_{\widetilde{\nu}^c}=k m^2_{\widetilde{N}}\, {\rm\,\,with\,\,}\,
k\in [-5.0,\,5.0]~.~ \quad
\end{align}
Additionally, we take $M_R=9m_D$ and $B_{M_R}=({100\rm\, GeV})M_R$,
and fix the irrelevant MSSM parameters as the following
\begin{align} \label{}
 &\tan\beta=20,\quad m_{\widetilde{l}}=10{\rm\,TeV},
 \quad \mu=400{\rm\,GeV},
\end{align}
while all the other sparticles are decoupled for simplicity.
Comments are in orders. First, the upper bound on $Y_\nu$ covers the
limit indicated by Eq.~(\ref{H:bound}) and is consistent with the
large Yukawa coupling of the third family. Second, we allow negative
$m_{\wt \nu^c}^2$ and $m_{\wt N}^2$ such that the LSP, via a
moderately large cancellation, can be light around $m_W$ even for
the large $Y_\nu$ and heavy $M_R$ case. Compared to the method
turning back to a large $B_{M_R}$ to get a light DM, this way allows
a widely varying mixing angle $\theta_{23}$. But they do not show
essential difference.

\subsubsection{Coannihilating Sneutrino LSP Inspired by the
Semi-Constrained ISS}

We now turn to a scenario inspired by the semi-constrained ISS-MSSM.
In this scenario, the MSSM part is described by the CMSSM with free
parameters $m_0$, $M_{1/2}$ and $A_0$, etc. The ISS-sector contains
the universal SUSY breaking soft mass terms $m_{\wt S}$, $A_\nu$ and $B_{M_R}$. For
the sake of enhancing Higgs boson mass, $|A_\nu|$ typically is
multi-TeV, says 5 TeV (But the enhancement is purely due to the
mixing effect and thus is limited, typically around 2 GeV.). Note
that $m_{\wt \nu_L}$, by naturalness, is favored to be at the
sub-TeV scale. So we need a rather small $Y_\nu$, typically at the order
of $0.001$, to control $C_{\wt \nu_L}$. The LSP annihilation rate is
suppressed by small $Y_\nu$, and then the coannihilation
effect~\cite{Griest} is needed to reduce its number density.

If the next-to-the LSP (NLSP) and the sneutrino LSP have
sufficiently small mass difference $\delta m$, coannihilation effect
will play an important role in reducing number density of the
sneutrino LSP. The effective annihilation cross section $\sigma_{\rm
eff}$, in terms of Ref.~\cite{Griest}, is a weighted sum of the
LSP-LSP, LSP-NLSP, and NLSP-NLSP annihilation cross sections, denoted
as $\sigma_{11}$, $\sigma_{12}$, and $\sigma_{22}$, respectively.
Viewing from the sneutrino system in the scenarios under
consideration, a large $\sigma_{\rm eff}$ should be ascribed to a
large $\sigma_{22}$ rather than $\sigma_{12}$.  The natural
candidates for the NLSP include the Higgsino and doublet-like
snerutrino which have full $SU(2)_L$ interactions. While the colored
sparticles, especially the light stop, should be moderately heavy
owing to the LHC bounds. So we do not consider them in this paper.

\begin{figure}[htb]
\includegraphics[width=3.2in]{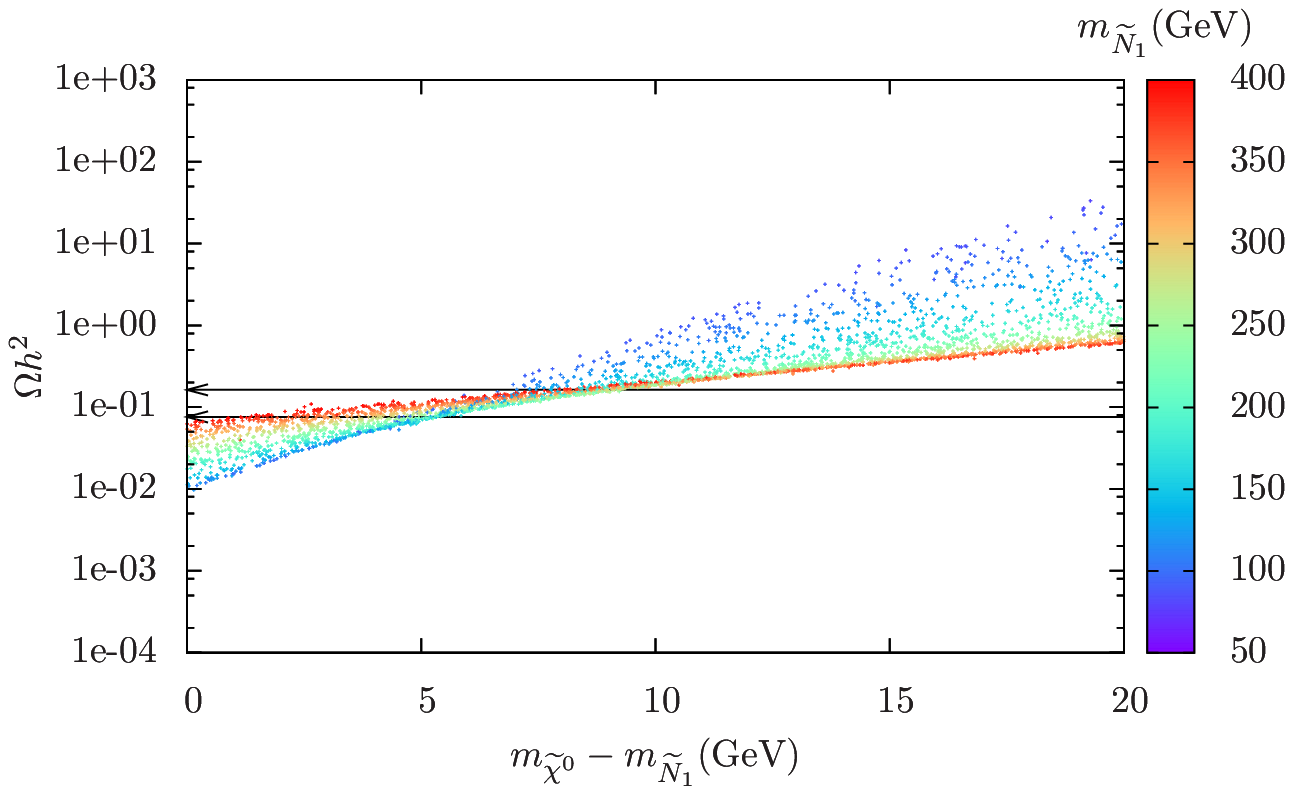}
\includegraphics[width=3.2in]{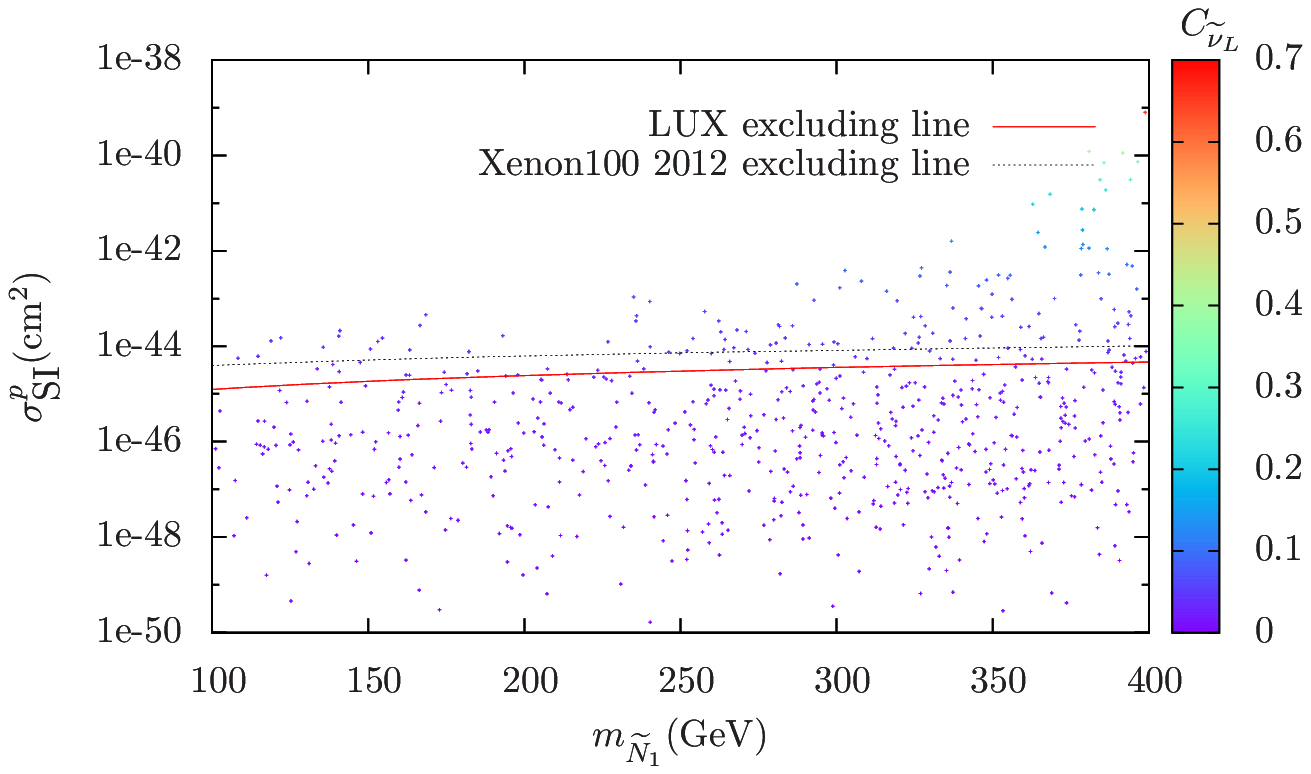}
\caption{Left: The DM relic density versus the  sneutrino-Higgsino
mass difference. Right: The direct detection exclusion limits on the sneutrino LSP 
in the $\sigma_{\rm SI}^p-m_{\rm DM}$ plane. The
heavier LSP may have a larger $C_{\wt \nu_L}$, so it is more
sensitive to direct detections.}\label{coa}
\end{figure}

We focus on the Higgsino NLSP case. This is a reasonable choice
because by naturalness the $\mu-$term should be small, and thus light
Higgsinos. We now make a numerical analysis. In terms of the above
arguments, we scan a slice of the parameter space as the following
\begin{align} \label{}
&A_\nu=5.0{\rm\,TeV},\quad Y_{\nu}\in [0.001,\,0.01], \quad\mu\in
[100,\,400]{\rm\,GeV} ,~\cr &
m_{\widetilde{\nu}^c}=m_{\widetilde{N}}\in [\mu-30,\,\mu+30],\quad
m_{\widetilde{l}}=[400,\,800]{\rm\,GeV}.
\end{align}
The other parameters are chosen as before. In this scenario the
bounds from direct detections are weak, see the right panel of
Fig.~\ref{coa} where most of the region is untouched, except for
those with a larger $C_{\wt\nu_L}$. This is not surprising. Because
$\sigma_{11}$ is allowed to be small,  the sneutrino LSP can
couple to Higgs, more widely, the visible sector, weakly. As a
consequence, it may become deeply dark, even for the indirect
detections. Instead, we may have to count on the possible hints if
 a light non-LSP Higgsino is discovered at colliders.

\section{Discussions and Conclusion}

The discovery of a relatively heavy SM-like Higgs boson is a good
news to SUSY but not to the MSSM, whose little hierarchy problem is
exacerbated. Moreover, the viable parameter space for the neutralino LSP
dark matter is greatly constrained. In this work we studied the two
aspects of the MSSM extended by the inverse seesaw mechanism, which
is an elegant mechanism to realize the low-scale seesaw mechanism without turning
to very small Yukawa couplings. This feature makes the model
contribute to a sizable radiative correction to $m_h$, up to $\sim$
4 GeV in the case of an IR-fixed point of the coupling 
$Y_\nu LH_u\nu^c$ and a large sneutrino mixing. Thus, the little hierarchy
problem can be alleviated. Furthermore, it makes the lightest
(highly complex) sneutrino be a viable thermal DM candidate. Owing
to the stringent constraints from the correct DM relic density and
XENON100 experiment, we found that there are only two viable scenarios
for the LSP sneutrino
\begin{itemize}
  \item Its dynamics is reduced to that of the Higgs-portal complex
  DM, with mass around $m_h/2$ or above $m_W$. The upcoming experiments
 such as XENON1T can exclude them, especially the latter. Additionally,
 we may observe a hint of DM with resonant enhancement from
 Higgs invisible decay.
  \item It is a coannihilating DM, likely with Higgsinos.
Because now a fairly weak coupling between the sneutrino DM and visible
particles is allowed, it is hard to be excluded.
\end{itemize}
Taking into account the requirement of enhancing $m_h$ which
needs large sneutrino mixing, we should introduce an extra family
of sneutrinos to account for sneutrino DM. And both 
scenarios naturally work when we attempted to suppress the DM
left-handed sneutrino component.

In this article we did not consider the supersymmetric ISS models
extended by gauge
groups~\cite{An:2011uq,Elsayed:2011de,Hirsch:2011hg}. Actually, such
supersymmetric models have even more significant effects to enhance
the SM-like Higgs boson mass~\cite{Elsayed:2011de,Hirsch:2011hg}.
The LHC detection in the presence of a singlet-like sneutrino DM
possesses some special signatures when the ordinary LSP is a stau
sneutrino~\cite{Mondal:2012jv}. Consequently, the colored sparticle
like stop decay is characterized by a long decay chain and the
presence of leptons in the final state, which may weaken the ATLAS
MET plus jets constraint. Other collider phenomeplogy
consequences of this model are also studied~\cite{Das:2013}.

\section*{Acknowledgement}

We would like to thank Tai Li and Valentina De Romeri for helpful
discussions. This research was supported in part by the China
Postdoctoral Science Foundation (No. 2012M521136) (ZK), by the Natural
Science Foundation of China under grant numbers 10821504, 11075194,
11135003, and 11275246 (JG, TL, YL), and by the National Basic Research 
Program of China (973 Program) under grant number 2010CB833000.

\end{document}